\documentclass[aps,prd]{revtex4}
\usepackage{subfigure,graphics} 
\usepackage{flafter} 
\begin{document} 

\title{
Particle production and reheating of the inflationary universe } 
\author{Ian G. Moss}
\email{ian.moss@ncl.ac.uk}
\author{Chris Graham}
\affiliation{School of Mathematics and Statistics, Newcastle University,
Newcastle Upon Tyne, NE1 7RU, UK}

\date{\today}


\begin{abstract}
Thermal field theory is applied to particle production rates in 
inflationary models, leading to new results for catalysed or two-stage 
decay, where massive fields act as decay 
channels for the production of light fields. 
A numerical investigation of the Boltzmann equation in an expanding universe shows that 
the particle distributions produced during small amplitude inflaton oscillations or 
even alongside slowly moving inflaton fields can thermalise.
\end{abstract}
\pacs{PACS number(s): }

\maketitle
\section{introduction}

Inflationary models give a picture of the early universe that has shown spectacular 
agreement with observation \cite{guth81,linde82,albrecht82}. 
All inflationary models require
a mechanism for reheating the universe to take it from the vacuum dominated 
inflationary phase to the 
hot radiation-dominated universe which we know must follow. 
The details of particular reheating mechanisms 
depend on the interactions between the inflaton and other fields, but 
the underlying process is particle production which fills the universe with
radiation.

The original work on reheating in the 1980's introduced particle production 
in an ad hoc fashion, assuming the rate of 
particle production by the inflaton $\phi$ in the limit of small $\dot\phi$ 
was proportional to $\dot{\phi}^{n}$ 
\cite{Albrecht:1982mp,Abbott:1982hn,turner83}. Most authors settled for
$n=2$, which has the advantage of being equivalent to a simple friction term in the inflaton
equation of motion. At around the same time, a less ad hoc approach was based on particle 
production caused by an inflaton field oscillating about the minimum of 
the inflaton potential after the end of inflation \cite{Dolgov:1982th,shafi84}. 
This latter approach appeared to be the more consistent, 
and it is still widely used today.

Following this early work, there where several attempts to apply new ideas in thermal 
field theory to the reheating process. These usually focused on finding effective field equations 
for the inflaton. The small-$\dot\phi$ equation of motion was derived first,
using linear response theory \cite{hosoya84,Morikawa:1984dz}. This has been used in the theory 
of warm inflation 
\cite{Moss85,yokoyama88,berera95}. 
More general forms of the inflaton field equation, which where not limited to small 
time derivatives and could be applied to the oscillating inflaton, followed later 
\cite{Boyanovsky:1993xf,gleiser94,Boyanovsky:1994me}. 

Renewed interest in particle production was ignited by the discovery of preheating, 
a nonperturbative process 
of inflaton decay though parametric resonance 
\cite{Traschen:1990sw,Kofman:1994rk,Shtanov:1994ce,Kofman:1997yn,Kofman:1997pt}. 
Like the earlier work, preheating 
involves particle production from an oscillating inflaton field. Preheating 
is a result of very large amplitude oscillations in the inflaton field. Large amplitude
oscillations are a feature, though not necessarily a desirable one, of most 
single-field inflationary models.

Preheating may or may not occur depending on the 
details of the particle model. 
In this paper we shall focus on models with a mechanism which we call catalysed or two-stage decay 
\cite{berera02}.  
Unlike in the case of pre-heating, we examine what happens when 
the fields coupled to the 
inflaton are too massive to be produced directly. Instead, the fields can act as decay 
channels for the production of light fields. This requires the kind of mass hierarchy which
is possible in supersymmetric models. Examples include Grand Unified extensions of the standard
model.

In section II we obtain formulae for the particle production rate due to changing 
particle masses or couplings and show that to a reasonable degree of accuracy 
the expansion of the universe can be neglected in making these calculations. In section III 
we consider several models of particle production that could arise during inflation, starting 
with the oscillating and slowly-evolving models, for which particle production rates are well-known 
and easily checked, and then we apply our method to the catalysed decay. 

In section IV we use the production rates calculated in the previous section to take a 
closer look at the evolution of a system which may be expanding and departing from thermal 
equilibrium. Thermalisation has previously been considered in the context of reheating through 
numerical solution of classical non-linear field theory \cite{Destri:2004ck,Micha:2004bv} or using 
a numerical solution for the non-equilibrium particle propagators \cite{Aarts:2007qu,Aarts:2007ye}.   
We solve the Boltzmann equation in an expanding universe 
with a source term  representing particle production. This allows us to 
consider models which would be too difficult to analyse directly using numerical 
approaches to quantum field theory.

\section{particle production}

In an ideal situation we would like to track the formation of particles during and after the 
inflationary era to give a full description of the reheating process. This requires a workable
definition 
of particle number. This definition need not be unique, but at least it should agree with 
the usual definition of particle number after inflation has ended. We shall use a definition 
of particle number which introduces a free field which instantaneously has the same field value and 
momentum as the interacting particle field.

A fundamental problem we face is that the particle production might not have a local description. 
However, we might hope that simple situations occur where the particles are produced at a 
rate depending only on local conditions, for example local field values or temperatures. 
For this reason, we focus on particle production rates.

We shall consider the production of particles due to a background time-dependent 
inflaton field $\phi$. The particles will be associated with a field $\sigma$.

\subsection{Particle number}

Following Morikawa and Sasaki \cite{Morikawa:1984dz}, we define particle creation and anihilation
operators for a
fiducial free field which coincides with the interacting field $\hat\sigma$ and momentum $\hat\pi$
at time $t$,
\begin{eqnarray}
\hat a^\dagger(p,t)&=&\omega_p\hat \sigma(-p,t)-i\hat\pi(p,t)\\
\hat a(p,t)&=&\omega_p\hat\sigma(p,t)+i\hat\pi(-p,t)
\end{eqnarray}
where $\omega_p=(p^2+m^2)^{1/2}$ may depend on time, and
\begin{equation}
\hat\sigma(p,t)=\int d^3x \,\hat\sigma(x,t)e^{-ip\cdot x}.
\end{equation}
The local number density is assumed to be spatially homogeneous. The number density in phase space
$n(p)$ can then be defined in terms of an ensemble average using the fiducial free field,
\begin{equation}
{1\over 2\omega_p}\langle\hat a^\dagger(p_1,t) \hat a(p_2,t)\rangle=
(2\pi)^3\delta(p_1-p_2)n(p_1).
\end{equation}
We shall express the density function in terms of the Wightman function $G_{21}(p,t_1,t_2)$, 
defined by
\begin{equation}
\langle \hat\sigma(p_1,t_1)\hat\sigma(p_2,t_2)\rangle=(2\pi)^3\delta(p_1-p_2)
G_{21}(p_1,t_1,t_2).
\end{equation}
The expression for the density function becomes
\begin{equation}
n(p,t)=\left[{1\over 2\omega_p}(\omega_p-i\partial_{t_1})
(\omega_p+i\partial_{t_2})G_{21}(p,t_1,t_2)\right],\label{neq}
\end{equation}
where $[\dots]$ will be used to indicate when a function is to be evaluated at $t_1=t_2=t$. The
frequency
$\omega_p$ always refers  to the value at time $t$, unless we state otherwise. 

\subsection{Particle creation}

In order to use Eq. (\ref{neq}), we would have to solve field equations for the Wightman 
function with suitable initial data, giving a particle density which is typically a non-local
function 
depending on the history of the background field. Instead of working directly with the density
function
directly, we shall look for a local approximation to the particle production rate. 

After some elementary manipulation, the particle production rate in phase space
obtained by taking the time derivative of Eq. (\ref{neq}) becomes
\begin{equation}
\dot n=\dot n_{\rm mass}+\dot n_{\rm int}.
\end{equation}
The first term represents particle production due to the changing particle
mass,
\begin{equation}
\dot n_{\rm mass}=\left[{\dot\omega_p\over 2\omega_p}
\left((\omega_p-i\partial_{t_1})+(\omega_p+i\partial_{t_2})
- {1\over\omega_p}(\omega_p-i\partial_{t_1})(\omega_p+i\partial_{t_2})
\right)G_{21}(p,t_1,t_2)\right]\label{nmass}
\end{equation}
The second term represents particle production at fixed particle mass,
\begin{equation}
\dot n_{\rm int}=-\left[{i\over2\omega_p}\left( 
(\partial_{t_1}^2+\omega_p^2)(\omega_p+i\partial_{t_2})
-(\omega_p-i\partial_{t_1})(\partial_{t_2}^2+\omega_p^2)
\right)G_{21}(p,t_1,t_2)\right]\label{nint}
\end{equation}

We can put these expressions into a more useful form by introducing the self-energy. Since we are
interested in the evolution of operators in a given initial state it proves convenient to use an
`in-in' formalism, and we use the Schwinger-Keldysh version \cite{schwinger61,keldysh64}. 
Propagators carry two extra internal
indices $a$ and $b$, where the indices $a$ and $b$ take the values $1$ or $2$. Following 
Calzetta and Hu \cite{calzetta88}, we raise and lower
indices with a metric $c_{ab}={\rm diag}(+1,-1)$.

The Schwinger-Dyson equations for the in-in formalism read \cite{calzetta88},
\begin{eqnarray}
\left(\partial_{t_2}^2+\omega_p^2(t_2)\right)G_{21}(p,t_1,t_2)&=&
-\int dt' G_2{}^a(p,t_1,t')\Sigma_{a1}(p,t',t_2)\label{sde1}\\
\left(\partial_{t_1}^2+\omega_p^2(t_1)\right)G_{21}(p,t_1,t_2)&=&
-\int dt' \Sigma_2{}^a(p,t_1,t')G_{a1}(p,t',t_2).\label{sde2}
\end{eqnarray}
For the terms in the particle production rate which have just one time 
derivative of the propagator, we use the LSZ trick of
introducing a time integral 
\begin{eqnarray}
\left.(\partial_{t_2}-i\omega_p)G_{21}(p,t_1,t_2)\right|_{t_2=t}&=&
\int_{-\infty}^t dt_2\,e^{-i\omega_p(t-t_2)}(\partial_{t_2}^2+\omega_p^2)
G_{21}(p,t_1,t_2)\label{it1}\\
\left.(\partial_{t_1}+i\omega_p)G_{21}(p,t_1,t_2)\right|_{t_1=t}&=&
\int_{-\infty}^t dt_1\,e^{-i\omega_p(t_1-t)}(\partial_{t_1}^2+\omega_p^2)
G_{21}(p,t_1,t_2).\label{it2}
\end{eqnarray}
We can now express the particle production rate in terms of integrals of the
propagator and the self-energy which have a suitable form for applying 
perturbation theory.

We shall consider the leading order in perturbation theory. First of all, let
\begin{equation}
\omega_p^2=p^2+m_\sigma^2+g^2\phi^2(t)
\end{equation}
where $m_\sigma$ is a constant and $\phi(t)$ is given. Suppose also that
$\Sigma=O(g^4)$. This is the kind of situation would arise, for example, given an inflaton $\phi$
and an interaction Lagrangian density ${\cal L}=g^2\phi^2\sigma^2/4$.

The leading order result, using Eq. (\ref{nmass}) and Eqs. (\ref{sde1}-{\ref{it2}), is that
\begin{equation}
\dot n_{\rm mass}={\rm Re}\left\{
{2g^2\phi\dot\phi\over \omega_p}\int_{-\infty}^t dt_2\,
{e^{-i\omega_p(t-t_2)}\over 2\omega_p}
(\phi^2(t)-\phi^2(t_2))G_{21}(p,t,t_2)\right\}
\end{equation}
where $G_{21}(p,t,t_2)$ is a free Wightman function for the $\sigma$-field
with the shifted mass (which need not be in the
vacuum state).

For the next term, we require the derivative of the Schwinger-Dyson equation (\ref{sde1}),
\begin{equation}
(\partial_{t_1}^2+\omega_p^2)(\partial_{t_2}^2+\omega_p^2)G_{21}(p,t_1,t_2)
=i\Sigma_{21}(p,t_1,t_2)+O(g^6).
\end{equation}
Use this together with Eqs. (\ref{nint}), (\ref{it1}) and (\ref{it2}),
\begin{equation}
\dot n_{\rm int}={\rm Im}\left\{
2\int_{-\infty}^t dt_2\,
{e^{-i\omega_p(t-t_2)}\over 2\omega_p}
\Sigma_{21}(p,t,t_2)\right\}.
\end{equation}
where $\Sigma_{21}$ is the self-energy of the $\sigma$-field at leading order.
We can see from this expression that $\dot n_{int}$ is the part of the production rate 
which is associated with the imaginary part of the self-energy.

\subsection{Curved space}\label{cs}

The formulae for the particle production rates found in the previous section neglected the expansion
of the universe. In this section we shall seek to show that expansion can be neglected to a
reasonable degree of accuracy when particle momenta are larger than the expansion rate.

Consider a spatially flat universe with scale factor $a$ and constant expansion rate $H$. 
Solutions to the wave equation in de Sitter space can be decomposed into particle modes
\cite{Birrell:1982ix} with
comoving wave number ${\bf k}$ which satsify
\begin{equation}
\left(\partial_{t}^2+3H\partial_{t}+\omega_p^2\right)f(k,t)=0
\end{equation}
where
\begin{equation}
\omega_p^2=k^2/a^2+m^2+12\xi H^2.
\end{equation}
A suitable normalisation is to use $f\dot f^*-f^*\dot f=i/a^3$. The modes can be expressed in terms
of Hankel functions,
\begin{equation}
f(k,t)={\sqrt{\pi}\over 2} Hk^{-3/2}z^{3/2}H^{(1)}_\nu(z),\label{csm}
\end{equation}
where $z=k/(Ha)$ and $\nu^2=9/4-m^2/H^2-12\xi$.

Consider a non-interacting field with Wightman function
\begin{equation}
G_{21}(k,t_1,t_2)=\left(N(k)+1\right)f(k,t_1)f^*(k,t_2)+N(k)f(k,t_2)f^*(k,t_1).\label{csw}
\end{equation}
The phase space number density is defined as before,
\begin{equation}
n(p,t)=\left[{a^3\over 2\omega_p}(\omega_p-i\partial_{t_1})
(\omega_p+i\partial_{t_2})G_{21}(k,t_1,t_2)\right].\label{ncseq}
\end{equation}
This is a function of the momentum ${\bf p}={\bf k}/a$ of a locally 
defined flat-spacetime theory. The factor $a^3$ is present because the 
Wightman function has undergone a shift in normalisation due
to the use of comoving modes. We substitute the Wightman function (\ref{csw}) 
and use the modes
(\ref{csm}). The result is quite complicated in general, but the 
main features can be seen in the
conformal case $m=0$ and $\xi=1/6$. In this case,
\begin{equation}
n=n_{deS}+n_{rad}.
\end{equation}
where
\begin{eqnarray}
n_{deS}\,\omega_p&=&\frac12{k\over a}\left(1+{3\over 2}{Ha\over k}\right)-\frac12\omega_p
\label{nds}\\
n_{rad}\,\omega_p&=&N{k\over a}\left(1+{3\over 2}{Ha\over k}\right)
\label{nN}
\end{eqnarray}
The non-vanishing contribution to the density function in the de Sitter vacuum is a 
reflection of
the fact that the particle number was defined using the physical momentum. 
The particle number
defined this way is analogous to the response of a particle detector. 
We can see from 
Eqs. (\ref{nds}) and (\ref{nN}) that $n\approx N$ for $p>>H$, so that 
in this limit we 
recover the flat space results. 

Similar considerations apply also to 
the particle
production rates. However, it is important to bear in mind when calculating particle 
production rates that $\dot n$ is evaluated at constant $p$ and not 
constant $k$,
\begin{equation}
\left({\partial n\over \partial t}\right)_p=
Hp\,\left({\partial n\over \partial p}\right)_t +\left({\partial n\over \partial t}\right)_k.
\label{ndotr}
\end{equation}
The first term on the right of this equation represents the reduction in particle density 
caused by the expansion of the universe. The second term, representing the particle production,
goes over to the flat space-time results when $p\gg H$. The redshift term is analysed
further in Sect. \ref{therm}.

It is interesting to integrate Eqs. (\ref{nds}) and (\ref{nN}) to get a formula for 
the energy density.
\begin{equation}
\rho_r=\rho_{deS}+\int {d^3 p\over (2\pi)^3}{k\over a}
N(k)\left(1+{3\over 2}{H^2a^2\over k^2}\right).
\end{equation}
The vacuum energy density of de Sitter $\rho_{deS}$ space comes from integrating 
the left hand side of Eq. (\ref{nds}), after we apply suitable 
regularisation methods. The full expression combines both curved space and thermal effects, 
with thermal effects dominating the integral for $p>>H$.

\section{examples}

We shall take a closer look at four different models and calculate some particle 
production rates using the formalism described in the previous section. These 
models are typical of what one might expect in
the context of inflation. Some of these results have been obtained before 
using other methods, and
these are included to check the consistency of the new approach.

\subsection{Oscillating fields}
\label{osc}

The first example we consider is the particle production from small amplitude 
oscillations of an
inflaton or other background field. In many inflationary models, this type of particle
production would be eclipsed by preheating from large
amplitude oscillations. However, this is not always the case, so that even this simple example may
be of interest.

The background field we take has
\begin{equation}
\phi=\phi_0(1+\epsilon\cos\, m_\phi t).\label{ib}
\end{equation}
where $\phi_0$ is the stable vacuum value of the field and $\epsilon$ is small 
variable which varies
slowly on the oscillation timescale. The field $\phi$ is coupled to a field $\sigma$ with effective
frequency $\omega_p$,
\begin{equation}
\omega_p^2=p^2+m_\sigma^2+g^2\phi^2-g^2\phi_0^2.\label{massterm}
\end{equation}
In this case we have introduced a shift so that the mass is $m_\sigma$ when $\phi=\phi_0$.

This particle production problem was solved long ago \cite{Dolgov:1998sf}. The total $2-$particle
production rate is given by the standard formula for particle decay,
\begin{equation}
\dot N={|{\cal M}|^2\over 8\pi}{p\over m_{\phi}}
\end{equation} 
where the momentum $p$ is determined by momentum conservation and the reduced matrix element 
${\cal M}$ is defined by
\begin{equation}
\langle {\bf p}_1,{\bf p}_2|0\rangle={\cal M}(2\pi)^4\delta({\bf p}_1+{\bf p}_2)
\delta(\omega_{p_1}+\omega_{p_2}-m_\phi)
\end{equation}
To leading order, the interaction with the background through Eq.  (\ref{massterm}) gives
\begin{equation}
\dot N={1\over 8\pi}g^4\phi_0^4\epsilon^2\left(1-{4m_\sigma^2\over m_\phi^2}\right)^{1/2}
\qquad m_\phi>2 m_\sigma.
\label{nscat}
\end{equation}
The dependence on the scalar field energy density $\rho_\phi=\phi_0^2\epsilon^2m_\phi^2$ is usually
factored to define the reheating coefficient $\Gamma$ by,
\begin{equation}
\Gamma={\dot\rho_r\over\rho_\phi}\approx{g^4\phi_0^2\over 8\pi m_\phi}.
\end{equation}
Since $g$ is typically very small, this type of perturbative reheating is quite inefficient and
would take several Hubble times to complete. 

We can consider the same problem, but using the general result for the time derivative of the
density function (\ref{nmass}). The free Wightman
function for the vacuum state is given by
\begin{equation}
G_{21}(t)={1\over 2\omega_p}e^{-i\omega_p t}.
\end{equation}
After integration over time, we have
\begin{equation}
\dot n={\pi\over 4}g^4\phi_0^4\epsilon^2{m_\phi\over \omega_p^3}\sin^2(m_\phi t)\,
\delta(\omega_p-m_\phi/2)
+{1\over 4}g^4\phi_0^4\epsilon^2{m_\phi^3\over \omega_p^4}
{\sin(2m_\phi t)\over (4\omega_p^2-m_\phi^2)}.
\end{equation}
The total particle creation rate integrated over momentum is
\begin{equation}
\dot N={1\over 4\pi}g^4\phi_0^4\epsilon^2\left(1-{4m_\sigma^2\over m_\phi^2}\right)^{1/2}
\sin^2(m_\phi t)
+{1\over 8\pi}g^4\phi_0^4\epsilon^2{m_\phi^2-8m_\sigma^2\over m_\phi m_\sigma}
\sin(2m_\phi t).
\end{equation}
This result appears complicated, but this is due to the presence of transient terms. Such terms
are to be expected when we try to calculate the particle production instantaneously. Over several 
oscillatory cycles, however, the production rate averages out
and we recover the scattering theory result (\ref{nscat}). The particle production in this 
case is effectively localised as long as we consider times longer than the oscillatory cycles.

\subsection{Derivative expansions}
\label{adi}

Another situation where we can have localised particle production rates is in the `adiabatic' limit,
when the inflaton has a small time derivative. This is a specialised form of particle production
which does not usually occur at leading order in perturbation theory. An important exception occurs
when the system starts out and remains close to thermal equilibrium. This type of particle
production was first discovered by Hosoya and Sakagama \cite{hosoya84} and by 
Morikawa and Sasaki \cite{Morikawa:1984dz}.

We start again from the general result for the particle creation rate (\ref{nmass}), using the
adiabatic approximation
\begin{equation}
\delta\phi^2(t_2)=\phi^2(t)-\phi^2(t_2)\approx 2\phi(t)\dot\phi(t)(t_2-t).
\end{equation}
We introduce transforms,
\begin{eqnarray}
G_{21}(t-t_2)&=&\int_{-\infty}^\infty{d\omega\over 2\pi}\,e^{-i\omega(t-t_2)}G_{21}(\omega)\\
\delta\phi^2(t_2)&=&
\int_{-\infty}^\infty{d\omega\over 2\pi}\,e^{i\omega(t-t_2)}\delta\phi^2(\omega)
\end{eqnarray}
After integration, we arrive at
\begin{equation}
\dot n=g^4\phi^2\dot\phi^2{G_{21}'(-\omega_p)\over \omega_p^2}.\label{npart}
\end{equation}
It only remains to give a formula for the thermal Wightman function. This can be expressed in terms
of a spectral function $\rho$ and the thermal distribution function $n$ 
(for example, see \cite{Berera:2008ar}),
\begin{equation}
G_{21}=-i(1+n)\rho.\label{g21}
\end{equation}
The spectral function typically has a Breit-Wigner form \cite{berera98},
\begin{equation}
\rho=(\omega^2-\omega_p^2-2i\omega\tau^{-1})^{-1}
-(\omega^2-\omega_p^2+2i\omega\tau^{-1})^{-1}\label{spec}
\end{equation}
where $\tau$ is known as the relaxation time. Inserting the Wightman function into Eq. (\ref{npart})
gives
\begin{equation}
\dot n=-g^4\phi^2\dot\phi^2{\tau n'(\omega_p)\over \omega_p^3}.
\end{equation}
This formula can also be derived using the methods of Ref. \cite{Morikawa:1984dz}, and it is closely
related to the work of Ref. \cite{hosoya84}. 

Note that the particle production is exponentially small for temperatures less than the
$\sigma-$particle
mass. In fact, Eq. (\ref{npart}) vanishes at zero temperature due to a general property of the
Wightman function. The best way to view this type of particle production is as a type of transport
phenomenon, similar to thermal or electrical conductivity. Particles are produced as the system
responds to the disturbance of thermal equilibrium caused by the changing mass. Increasing the
relaxation time $\tau$ gives more time for the mass to change, driving the system further from
equilibrium and
increasing the particle production.

The particle production takes energy from the inflaton field, and we can ensure energy balance by
introducing a friction term $\Gamma\dot\phi$ into the inflaton equation. The total radiation energy
is
\begin{equation}
\rho_r=\int {d^3p\over (2\pi)^3}\,n\omega_p
\end{equation}
The time variation of $\rho_r$ contains a term from the time variation of the $\omega_p$, which
relates to the change with time of the inflaton effective potential, and a $\Gamma\dot\phi^2$ term,
where
\begin{equation}
\Gamma={\dot\rho_r\over \dot\phi^2}=
-g^4\phi^2\int {d^3p\over (2\pi)^3}{\tau n'(\omega_p)\over \omega_p^2}.
\end{equation}
The friction coefficient obtained from the particle production formula agrees with the friction
coefficient obtained  in the inflaton equation of motion using linear response theory
\cite{hosoya84}.

\subsection{Catalysed or two-stage decay}
\label{catosc}

In the third model of particle creation an oscillating inflaton decays into light thermal scalar
particles through an intermediate virtual boson. This is a natural set-up, because many particle
models contain both heavy and light fields, with the light particle masses protected by
supersymmetry. Particles which couple to the inflaton will tend to be
massive, and may well be too heavy to be produced directly by the inflaton oscillations. Preheating
is supressed, and we have to rely on perturbative particle production effects. 
We shall suppose that the model is part of a supersymmetric theory, which protects the 
flatness of the inflaton potential
and the masses of the light fields, doing away with the need to fine-tune their coupling constants.

The mass of the light particles in this example is independent of time, and we use the second
formula (\ref{nint}) for the particle production rate. Having a fixed mass removes some of the
ambiguities in the definition of the particle number, and
leads to a `cleaner' result.

The heavy field is denoted by $\chi$ and the light field by $\sigma$. The interaction Lagrangian we
shall take is
\begin{equation}
{\cal L}_I=-\frac14 g^2\phi^2\chi^2-\frac12h\,m\,\sigma^2\chi-\frac1{4!}\lambda\sigma^4,\label{lint}
\end{equation}
where $g$, $h$, $\lambda$ and $m$ are constants. We can choose $m=g\phi_0$ by redefining $h$ if
necessary. The self-interaction term of the light fields is
included to allow them to come to thermal equilibrium. We shall consider particle production
into the vacuum and also in the presence of thermal radiation. 

\begin{center}
\begin{figure}[ht]
\scalebox{0.5}{\includegraphics{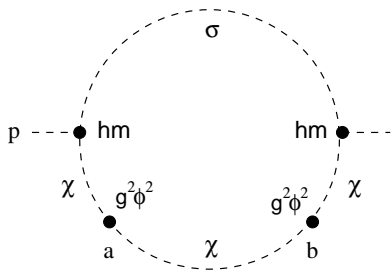}}
\caption{The Feynman diagram contributing to the imaginary part of the $\sigma$ self energy
 $\Sigma_{12}$.}
\label{figsigma}
\end{figure}
\end{center}

The background is as before Eq. (\ref{ib}), but now $m_\chi>m_\phi$. The first non-trivial
contribution to the imaginary part self-energy is given by the Feynman diagram show in figure
\ref{figsigma}. We define the fourrier transform as in the previous section and
then the diagram contributes
\begin{eqnarray}
\Sigma_{21}(p,t,t_2)&=&
g^4 h^2 m^2\int {d^3k\over (2\pi)^3}{d\omega_1\over 2\pi}{d\omega_2\over 2\pi}
{d\omega_3\over 2\pi}\,e^{-i\omega_2(t-t_2)}G_{\sigma21}({\bf p}-{\bf k},t-t_2)\nonumber\\
&&G_{\chi2}{}^a({\bf k},\omega_1)\phi^2(\omega_1-\omega_3)
G_{\chi a}{}^b({\bf k},\omega_3)\phi^2(\omega_3-\omega_2)
G_{\chi b1}({\bf k},\omega_1),\label{sigma}
\end{eqnarray}
where a subscript has been used to distinguish between $\chi$ and $\sigma$ propagators.

We concentrate on the low energy spectrum $p<<m_\chi$, when we can use a low momentum approximation
for the $\chi$ propagators,
\begin{eqnarray}
G_{\chi2}{}^2({\bf k},\omega)&\approx&-{i\over m_\chi^2}\\
G_{\chi2}{}^1({\bf k},\omega)&\approx&{\alpha\over m_\chi^4}\theta(\omega)\\
G_{\chi11}({\bf k},\omega)&\approx&-{i\over m_\chi^2}
\end{eqnarray}
where $\theta(\omega)$ is the heaviside function. The middle equation follows from Eqs. (\ref{g21})
and (\ref{spec}), where  
$\tau^{-1}$ is now the heavy particle decay width and 
$\alpha=4\omega_p/\tau\propto h^2 m^2$. We use the free Wightman function
for the $\sigma$ field with occupation number $n$.

With these approximations, using Eq. (\ref{nint}) for the particle production rate,
\begin{equation}
\dot n_{int}={g^2h^2\over 2\pi^2}{m^4m_\phi^2\phi_0^2\epsilon^2\over \tau m_\chi^8}
 F(p),\label{ndotC}
\end{equation}
where
\begin{equation}
F(p)=
\int {k^2 dk\over \omega_k m_\phi^2}
\theta(m_\phi-\omega_p-\omega_k)\left(1+n(\omega_k)\right).
\end{equation}
For small $m_\sigma$, the integral gives a dilogarithm function,
\begin{equation}
F(p)={T^2\over m_\phi^2}{\rm dilog}\left(e^{(m-\omega_p)/T}\right).
\end{equation}
The function $F(p)$ has been plotted in figure \ref{figsm}. As might be expected, the vacuum
particle production rate peaks when the momentum equals half the inflaton mass. The physical
process behind the particle production this time is a decay $\phi\to 4\sigma$, using two
intermediate virtual $\chi$ bosons. The four particle decay is reflected in the broad width of the
peak, compared to the resonance in the $2-$particle decay in Sect. \ref{osc}.

\begin{center}
\begin{figure}[ht]
\scalebox{0.7}{\includegraphics{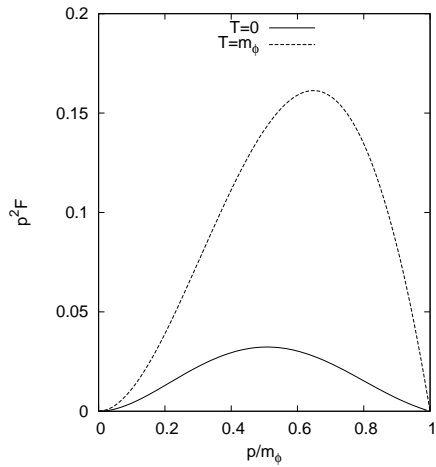}}
\caption{The momentum dependence of the particle production rate for the two-stage decay with an
oscillating inflaton (model C). The function $p^2F(p)$
has been plotted for $m_\sigma=0$ at $T=0$ and $T=m_\phi$.}
\label{figsm}
\end{figure}
\end{center}

In the zero temperature limit, the reheating coefficient $\Gamma$ which describes the rate of
production of radiation becomes
\begin{equation}
\Gamma={\dot\rho_r\over \rho_\phi}=
{\alpha g^2h^2\over 480\pi^4}{m^4m_\phi^3\over m_\chi^8}.\label{gamma}
\end{equation}
For $\alpha\sim h^2m^2$, this is smaller than the corresponding result in Sect. \ref{osc} for the
$m_\phi>m_\chi$ regime. However, in supersymmetric models, the couplings do 
not have to be especially small. Furthermore, if there are many species of light 
scalar fields, then the
the final result scales with the number of fields. 

\subsection{Derivative expansion for catalysed or two-stage decay}
\label{cat}

The final example is another `adiabatic' process, but this time the low temperature
behaviour is suppressed by a power law instead of the exponential suppression found in Sect.
\ref{adi}. The inflaton decays into light thermal scalar particles through an
intermediate virtual boson as in the previous example. This model was introduced in the context of
warm inflation  \cite{berera02}, but the set-up can occur quite easily in models which contain both
heavy and light particles.

The interaction Lagrangian is the one used in the previous section, Eq. (\ref{lint}).
We take an initial state to be one of thermal radiation with temperature $T<<m_\chi$.
How the system might come to thermal equilibrium will be addressed in the next section.

The first non-trivial contribution to the self-energy is again given by the Feynman diagram show in
figure \ref{figsigma} and the expression (\ref{sigma}). We can use the condition $T<<m_\chi$ to
justify a low momentum approximation for
the $\chi$ propagators again, now with
\begin{eqnarray}
G_{\chi2}{}^2({\bf k},\omega)&\approx&-{i\over m_\chi^2}\\
G_{\chi2}{}^1({\bf k},\omega)&\approx&{\alpha\over m_\chi^4}(1+n)\\
G_{\chi11}({\bf k},\omega)&\approx&-{i\over m_\chi^2}
\end{eqnarray}
where the middle equation follows from Eqs. (\ref{g21}) and (\ref{spec}), with 
$\alpha=4\omega/\tau$. In general, $\alpha$ is a function of momentum and energy, which has been
calculated explicitly for non-zero temperatures in Ref. \cite{Moss:2006gt}.

We use an adiabatic approximation for the inflaton fields as in Sect. \ref{adi},
\begin{equation}
\delta\phi^2(\omega)=2i\phi\dot\phi(2\pi \delta'(\omega))
\end{equation}
where the primes denote derivatives with respect to $\omega$. With these approximations, the
self-energy becomes
\begin{equation}
\Sigma_{21}(p,t,t_2)=
4g^4 h^2 {m^2\phi^2\over m_\chi^8}\dot\phi^2\int {d^3k\over (2\pi)^3}{d\omega\over 2\pi}
\,e^{-i\omega(t-t_2)}G_{\sigma21}({\bf p}-{\bf k},t-t_2)
\left(\alpha(\omega)[1+n(\omega)]\right)^{\prime\prime}.
\end{equation}
Now we can apply formula (\ref{nint}) for the particle production rate, using the free thermal
propagator for the $\sigma$ field,
\begin{equation}
\dot n_{int}=-4g^4 h^2 {m^2\phi^2\over \omega_p m_\chi^8}\dot\phi^2\int {d^3k\over (2\pi)^3}
{\alpha\over \omega_k}\left\{ n''(-\omega_k-\omega_p)(1+n(\omega_k))\right\}.
\end{equation}
So far, we have not had to assume a particular form for the distribution function $n$. However, if
$n$ is the thermal distribution function, then the integral can be done approximately in the small
$m_\sigma$ mass limit,
\begin{equation}
\dot n_{int}={g^2 h^2\over 2\pi^2} {m^4\dot\phi^2\over \tau m_\chi^8}F(p)
\label{ndotD}
\end{equation}
where $m=g\phi$ and
\begin{equation}
F(p)=T^2
\left\{ n(\omega_p)\sum_{n=1}^\infty{1\over n^2}\left(1-e^{-n\omega_p/T}\right)\right\}''.
\end{equation}
The function $F(p)$ has been plotted in figure \ref{figsp}.
Note that the pre-factor is identical with the pre-factor in Eq. (\ref{ndotC}) for the oscillating
case if we use the period averaged value of $\dot\phi^2$. The momentum distribution is quite
different, however, and the particle production vanishes as $T\to 0$. 

\begin{center}
\begin{figure}[ht]
\scalebox{0.7}{\includegraphics{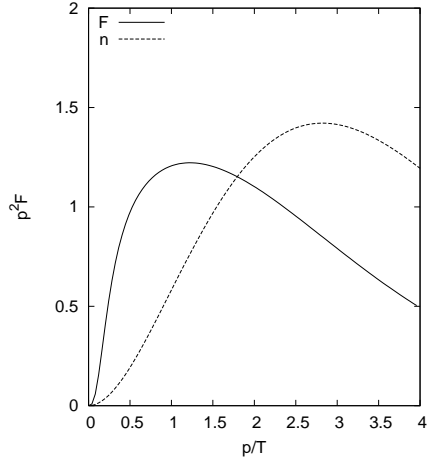}}
\caption{The momentum dependence of the particle production rate for small $\dot\phi$ with the
two-stage decay (model D). The function $p^2F(p)$ 
is plotted. The thermal distribution $p^2n$ is shown for comparison.}
\label{figsp}
\end{figure}
\end{center}

The reheating coefficient $\Gamma$ can be obtained by integrating over momentum,
using $\alpha=4\omega_p/\tau\approx const$,
\begin{equation}
\Gamma={\dot\rho_r\over \dot\phi^2}={\alpha g^2 h^2\over 4\pi^4}{m^4\over m_\chi^8}T^3
\label{gammat3}
\end{equation}
This agrees with the friction coefficient calculated from the effective field equations in ref.
\cite{Moss:2006gt}.

\section{thermalisation}
\label{therm}

The adiabatic particle production rates where calculated under the assumption that the system was
close the thermal equilibrium. In this section we shall examine the validity of this assumption by
solving the Boltzmann equation in an expanding universe. This will also give us an opportunity to
consider systems which
depart from equilibrium.

We shall adopt a pseudo-particle approximation where the Wightman function take a thermal form, but
with an arbitrary distribution function $n\equiv n(p,t)$. The particle number will evolve according
to
\begin{equation}
\dot n={\cal S}_p+{\cal S}_r+{\cal S}_c.\label{evn}
\end{equation}
where ${\cal S}_p$ represents particle production, ${\cal S}_r$ represents particle dilution due to
the expansion of the universe and ${\cal S}_c$ is the Boltzmann collision term

The particle production rates calculated in Sects. \ref{adi}--\ref{cat} are still valid with
the new distribution functions and can be used for ${\cal S}_p$. The expansion effect 
represents a stretching of the physical
wavelengths of the modes by the scale factor $a$. This term was evaluated in Eq (\ref{ndotr}),  
\begin{equation}
{\cal S}_r=Hp\,\partial_p n.
\end{equation}
The collision term for $2\to2$ particle scattering from the quadratic
term in the Lagrangian density Eq. (\ref{lint}) is
\begin{equation}
{\cal S}_c={\lambda^2\over 2\omega_p}\int {d^3p_2\over2\omega_{p_2}}
{d^3p_3\over2\omega_{p_3}}{d^3p_4\over2\omega_{p_4}}
(2\pi)^{-5}
\delta(P+P_2-P_3-P_4)
B(p,p_1,p_2,p_3),\label{scol}
\end{equation}
where $P=(\omega_p,{\bf p})$ and
\begin{equation}
B(p,p_1,p_2,p_3)=(1+n(p))(1+n(p_2))n(p_3)n(p_4)-n(p)n(p_2)(1+n(p_3))(1+n(p_4)).
\end{equation}
This term preserves the total 
particle number as well as the total energy. 

Multiplying the Boltzmann equation by $\omega_p$ and integrating gives the total energy equation
\begin{equation}
\dot\rho_r+4H\rho_r=S,
\end{equation}
where the source term is
\begin{equation}
S=\int{d^3 p\over (2\pi)^3}{\cal S}_p\omega_p.
\end{equation}
In the oscillating inflaton case $S=\Gamma\rho_\phi$ and in the slowly evolving inflaton limit 
$S=\Gamma\dot\phi^2$, where expression have been given for $\Gamma$ in Sects. \ref{adi}--\ref{cat}. 
Both types of reheating coefficient can be combined into
\begin{equation}
\dot\rho_r+4H\rho_r=\Gamma(\rho_\phi+p_\phi),\label{energycons}
\end{equation}
where $p_\phi$ is the  averaged pressure term. Various forms of this equation have been used
in the past to study reheating \cite{turner83} and warm inflation \cite{berera97}.

The simplest way to analyse the Boltzmann equation (\ref{evn}) is to take a close-to-equilibrium
approximation, introducing a thermal distribution function $n_T$ and defining the effective
temperature by
matching the energy density with the actual distribution function,
\begin{equation}
\int {d^3 p\over (2\pi)^3}(n-n_T)\omega_p=0.
\end{equation}
We can use the thermal particle production rates calculated earlier and introduce a 
thermal relaxation-time $\tau_r$ to simplify the collision term. The
Boltzmann equation we have to solve is then
\begin{equation}
\dot n=RF(p)+Hp\,\partial_p n-\tau_r^{-1}(n-n_T),\label{brt}
\end{equation}
where $R$ is the prefactor in the particle production rates Eq. (\ref{ndotC}) or Eq. (\ref{ndotD}). 
We have 
integrated this equation numerically using a fourth order
Runge-Kutta scheme for the time derivatives and second order differences for the momentum
derivatives. This numerical procedure is very fast and stable, with most of the results given below
taking less than one second on a 1GHz processor.

\subsection{Oscillating phase}

We consider a period of reheating with an oscillating inflaton and
$\rho_\phi>>\rho_r$. This regime ends, according to Eq. (\ref{energycons}), when $H\approx\Gamma$. 
During
this period, the pressure averages to zero over the oscillation period and the universe expands
like a pressure free cosmological model, with
\begin{eqnarray}
H(t)&=&H(0)\left(1+\frac32 H(0)t\right)^{-1},\\
R(t)&=&R(0)\left(1+\frac32 H(0)t\right)^{-2}.
\end{eqnarray}
The second equation follows from $R\propto \rho_\phi$.

Some numerical results for the momentum distribution obtained from Eq. (\ref{brt}) are shown in
figure \ref{figbolzC}. The distribution thermalises, and does so more quickly with smaller
relaxation times as might be expected. The momentum distribution of the source term shows up
clearly at early times, before the relaxation has taken effect.

\begin{center}
\begin{figure}[ht]
\scalebox{0.5}{\includegraphics{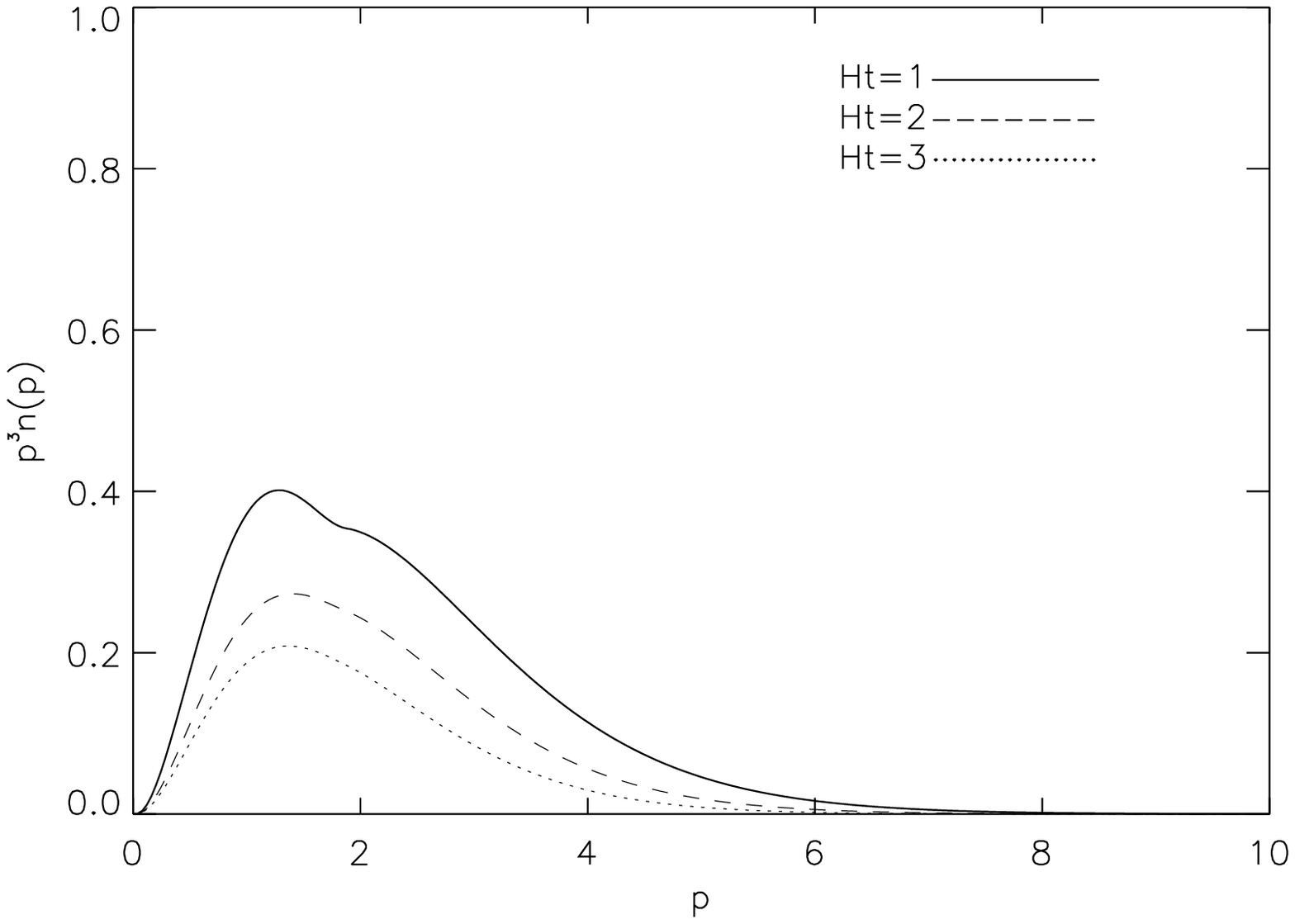}}
\scalebox{0.5}{\includegraphics{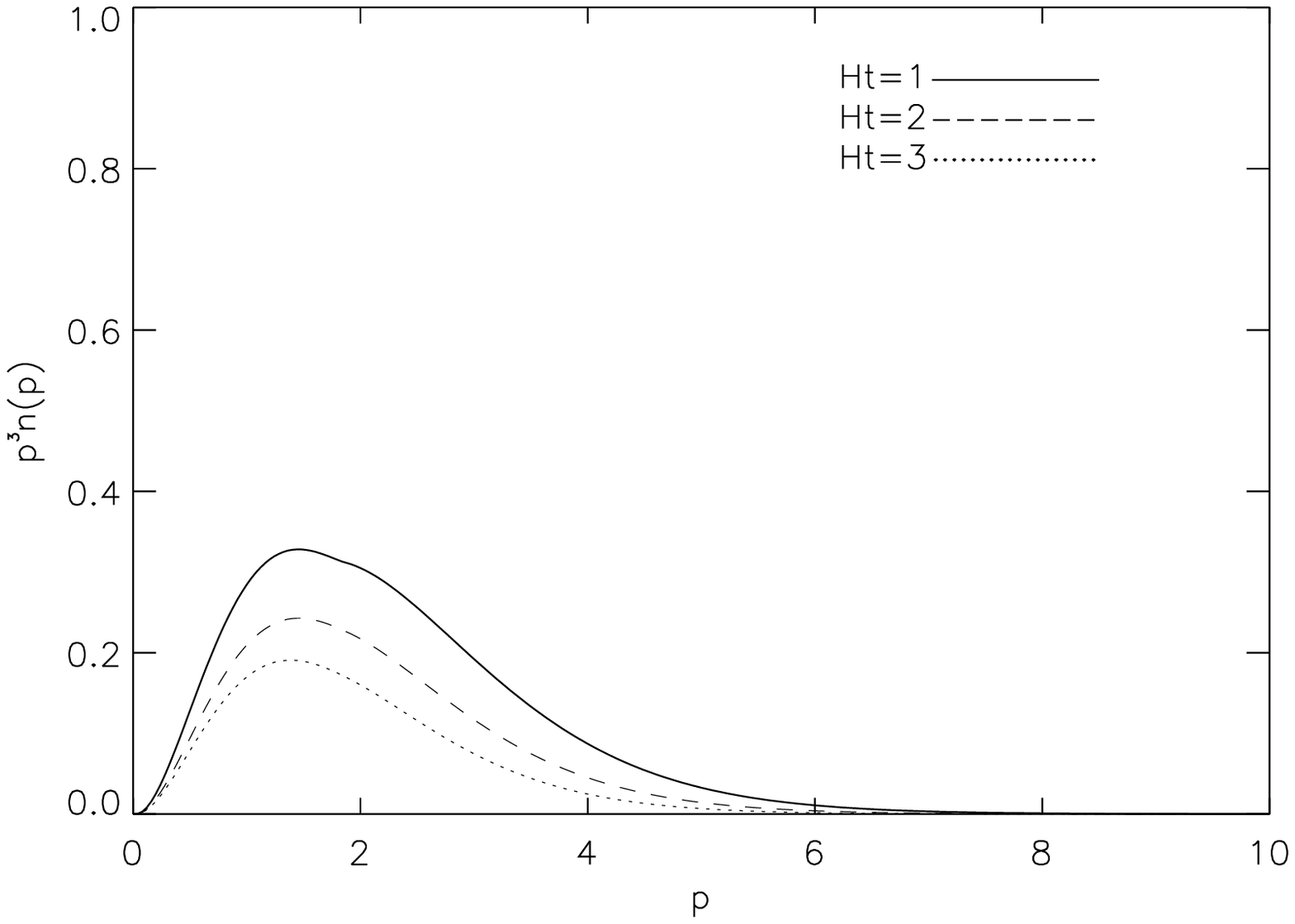}}
\caption{The stationary momentum distribution for the oscillatory phase in the two-stage decay
(model D) using the relaxation-time approximation. The relaxation times 
are $\tau_r=0.1/H(0)$ (left) and $\tau_r=0.05/H(0)$ (right). 
As might be expected, shorter relaxation
times produce a spectrum which is closer to thermal equilibrium. The constant 
$R(0)=50H(0)$ and
$m_\phi=2H(0)$.}
\label{figbolzC}
\end{figure}
\end{center}

\begin{center}
\begin{figure}[ht]
\scalebox{0.5}{\includegraphics{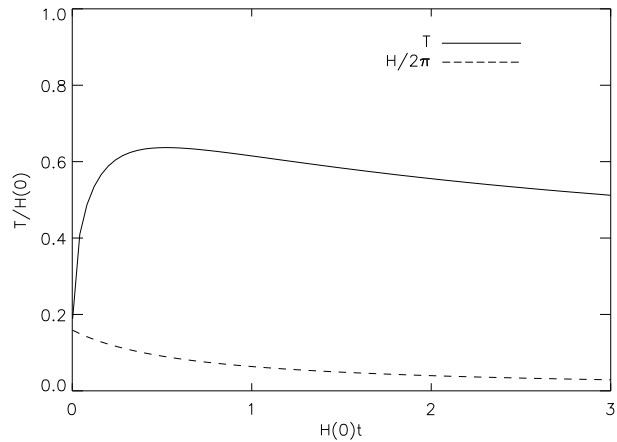}}
\caption{The time evolution of the effective temperature for the oscillatory phase with the
two-stage decay (model D) using the relaxation-time approximation. The de Sitter temperature
$H/2\pi$ is shown for comparison. The relaxation time 
$\tau_r=0.1/H(0)$,  the constant $R(0)=50H(0)$ and $m_\phi=2H(0)$.}
\label{figtimeC}
\end{figure}
\end{center}

The initial temperature for the numerical solutions has been set equal to the de Sitter temperature
$H(0)/2\pi$, to be consistent with the assumptions used in the particle production calculations.
The evolution of the temperature is shown in figure \ref{figtimeC}. After a sharp rise to a
maximum, the temperature falls off as $t^{-1/4}$. This agrees very well with the analytic solution
to the total energy equation (\ref{energycons}) \cite{turner83,Chung:1998rq}.

\subsection{Slow-roll phase}

Small values of $\dot\phi$ are characteristic of the slow-roll phase of inflation. The particle
production rates calculated in sect \ref{cat} can be applied to the slow-roll phase, 
provided we can justify the thermal
hypothesis which was used. During the slow-roll phase of inflation, both $H$ and $\Gamma$ vary 
very little over several Hubble
times, and we can treat them as constants in the Boltzmann equation (\ref{brt}). 

\begin{center}
\begin{figure}[ht]
\scalebox{0.5}{\includegraphics{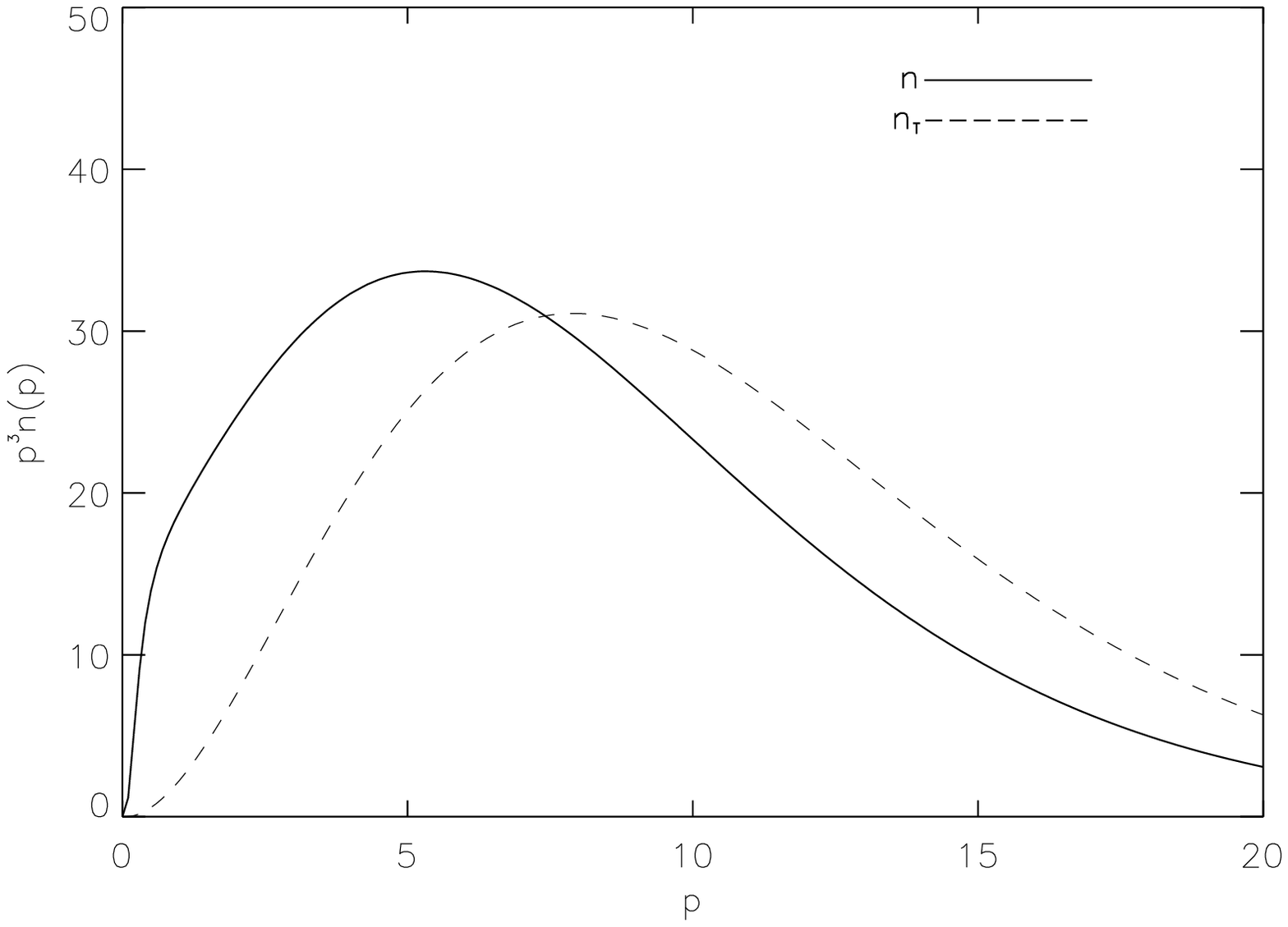}}
\scalebox{0.5}{\includegraphics{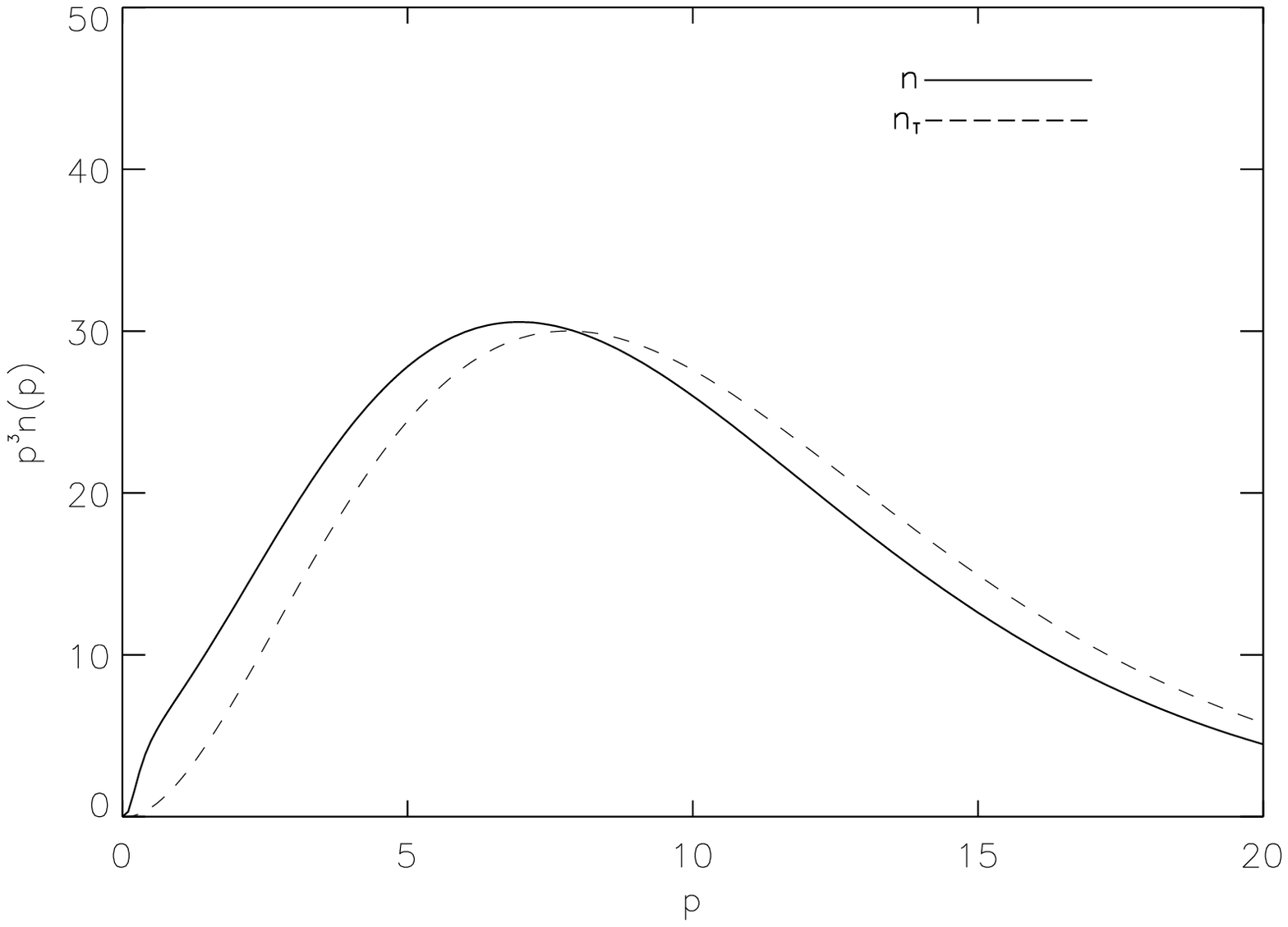}}
\caption{The stationary momentum distribution for different relaxation times in the
two-stage decay (model D) using the relaxation-time approximation. The relaxation times 
are $\tau_r=0.2/H$ (left) and $\tau_r=0.05/H$ (right). As might be expected, shorter relaxation
times produce a spectrum which is closer to thermal equilibrium. The constant $R=15H$ and
$m_\sigma=0.25H$.}
\label{figbolz}
\end{figure}
\end{center}

\begin{center}
\begin{figure}[ht]
\scalebox{0.5}{\includegraphics{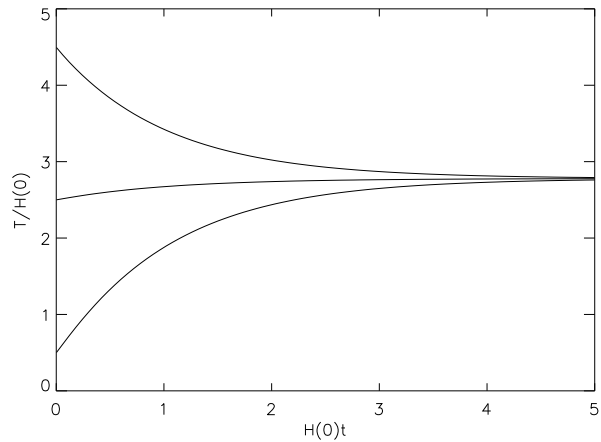}}
\caption{The time evolution of the effective temperature for different initial conditions with the
two-stage decay (model D) using the relaxation-time approximation. In each case, the momentum
distribution reaches a stationary state with the effective temperature shown. The relaxation time 
$\tau_r=0.1/H$,  the constant $R=15H$ and $m_\sigma=0.25H$.}
\label{figtime}
\end{figure}
\end{center}

Numerical solutions for two different parameter sets are shown in Fig. \ref{figbolz}. These
show the existence of an attractor with non-zero temperature and a spectrum close to
thermal equilibrium. The final momentum distribution does not show any dependence on the initial
distribution, but it is
dependent on the relaxation time, a shown in figure \ref{figbolz}. Small relaxation times,
corresponding to relatively large values of the self-coupling $\lambda$, lead to nearly thermal
spectra.

The parameters for the numerical solution where chosen to place the temperature in the range $T>H$
required for consistency of the particle production calculations. The time evolution of the
temperature shown in figure \ref{figtime} agrees very well with the analytic solution to 
Eq. (\ref{energycons}) when 
$\Gamma\propto T^3$ (see Eq. (\ref{gammat3})), which has the
form
\begin{equation}
T=T_\infty\left(1-e^{-Ht}\right).
\end{equation}
This shows clearly how the expansion of the inflationary universe need not lead to a 
supercooled state when particle production is taken into account.

\subsection{Thermalisation with the full boltzmann collision integral}

In the above work we have introduced the thermal relaxation-time $\tau_r$ to approximate the 
thermalisation effects of the Boltzmann collision term. We can check the validity of this
approximation by solving the the Bolzman equation with the full $2\to2$ particle scattering term
(\ref{scol}). Following the work of Refs. \cite{Dolgov:1998sf,Bijlsma:2000}, we can eliminate the
delta-functions and reduce the integral from 9 to 2 dimensions, which gives
\begin{equation}
{\cal S}_c=
{D\over \omega_p p}\int \theta(\omega_{p_2}-m_\sigma)
{\rm min}(p,p_2,p_3,p_4)B(p,p_2,p_3,p_4)d\omega_{p_3} d\omega_{p_4},
\end{equation}
where $D=\lambda^2/64\pi^3$ and 
$\omega_{p_2}\equiv \omega_{p_2}(p,\omega_{p_3},\omega_{p_4})$ is obtained from energy
conservation,
\begin{equation}
\omega_{p_2}=\omega_{p_3}+\omega_{p_4}-\omega_p.
\end{equation}

We have solved equation (\ref{brt}) numerically with the new expression for ${\cal S}_c$, focussing 
on the two stage decay model (model D). Again, we used a fourth order Runge-Kutta scheme for time 
derivatives and second order differences for the momentum derivatives. The collision term
was evaluated using a 2D Simpson's rule integrator. In order to remove instability problems at low
momenta we damped the source term with a factor $p^2/(p^2+H^2)$, which is consistent with our
calculation of the source term which cannot be used for $p$ less than $H$. We also avoided using a
very
fine momentum mesh that would bring in grid points at very low momentum. Solving with the full
collision
term is computationally far more demanding than using a relaxation-time approximation. For
reasonable
mesh sizes the total integration times are approximately an hour on GHz processors, compared to
one second for the relaxation-time approximation.   

Numerical results for the full collision term with two stage decay model are shown in figures 
\ref{figbolzfull} and \ref{figtimefull}, obtained using the same values for constants 
$R$ and $m_\sigma$ as before. The distribution reaches a stable non-zero temperature as expected and
is  consistent with the findings using the relaxation-time approach. 

Comparison of Fig. \ref{figbolz} and Fig. \ref{figbolzfull} suggests that the relaxation-time 
$\tau_r=0.1H^{-1}$ corresponds to $D\approx10$. This example is strongly self-coupled. 
However, it is possible to argue that value of $D$ needed for thermalisation decreases 
if we increase the particle production rate. According to dimensional analysis, 
the relaxation time should be proportional to the inverse temperature. 
The numerical example has $T=2H$, hence $D\approx2/(T\tau_r)$. We therefore predict
a similar distribution function to Fig. \ref{figbolzfull}  for $D<1$ when the particle production is
increased to give an effective temperature $T>20H$.

\begin{center}
\begin{figure}[ht]
\scalebox{0.5}{\includegraphics{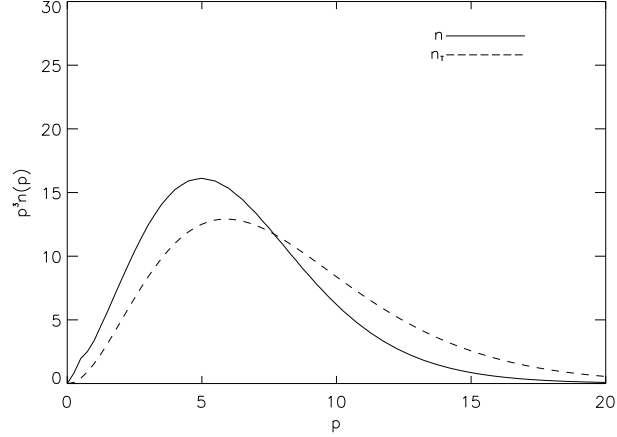}}
\caption{The stationary momentum distribution in the two-stage decay (model D) using 
the full collision integral gives a check for consistency of the relaxation time approximation
used in Fig. \ref{figbolz}. The 
parameters are $R=15H$ and $m_\sigma=0.25H$ and $D=10$. The plot is
comparable to Fig. \ref{figbolz} with a relaxation time $\tau_r=0.1/H$.}
\label{figbolzfull}
\end{figure}
\end{center}

\begin{center}
\begin{figure}[ht]
\scalebox{0.5}{\includegraphics{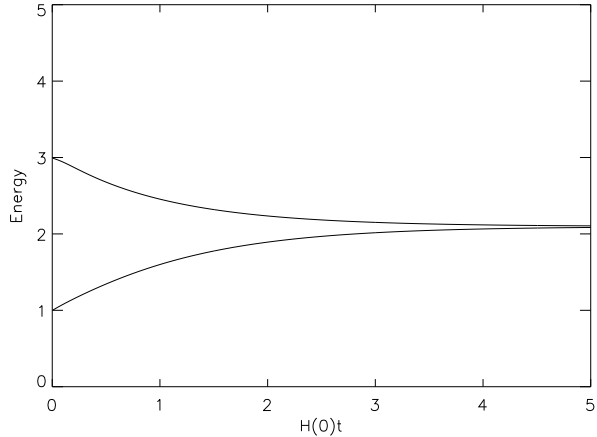}}
\caption{The time evolution of the effective temperature for different initial conditions with 
the two-stage decay (model D) using the full collision integral. As with the relaxation time 
approximation, the momentum distribution reaches a stationary state with the effective 
temperature shown. The constant $D=10$, with $R=15H$ and $m_\sigma=0.25H$.}
\label{figtimefull}
\end{figure}
\end{center}

\section{conclusion}
We have attempted to produce a uniform description of particle production during the early universe
which can cope with oscillating and slowly varying inflaton background fields. We have concentrated
mainly on a two-stage decay process where the inflaton decays into light radiation fields through
an intermediate heavy boson. 

Thermalisation of the particles has been described by solving the Boltzmann equation in an expanding
universe. We have found that the thermisation and particle production can be combined to produce a
prediction for the momentum distribution in the radiation fields. In many cases, where the
self-interactions allow, the distribution approaches a thermal distribution. 

Our results for particle creation and thermalisation in the case of a slowly-evolving inflaton 
field are fully consistent with the thermal dissipation processes predicted in warm inflation
\cite{berera98,Moss:2006gt}. Most of these models have have assumed that the radiation 
remains close to thermal equilibrium, and we have found that this occurs when the the 
self-coulping of the radiation field is sufficiently large.

The particle production rates, and therefore the thermal dissipation rates, are still significant 
even when the the distribution function departs substatially from thermal equilibrium. Distortions 
to the spectra due to the finite relaxation time of the radiation may have observational 
consequences if the thermal fluctuations are the source of density fluctuations 
in the cosmic microwave background \cite{Moss85,bererafang95,taylor00,Hall:2003zp}. Further work
would be 
worthwhile to find the effect this may have on the spectrum and as a source of non-gaussianity.

The reason for considering the two-stage decay lay partly in the fact that there are light fields
whose masses are protected by supersymmetry. In a supersymmetric model, the bosonic decays which we
have considered would be accompanied by fermionic decays. The extension of
the present results to fermions is tedious, but straightforward. Fortunately, fermion decays tend to
be
suppressed at low temperatures, compared to the bosonic ones \cite{Moss:2006gt}, and so it 
should be reasonable to ignore them.

An interesting regime occurs when the temperature is comparable to the expansion rate. 
Both the thermal equilibrium and flat-spacetime approximations break down in this limit. We have
suggested ways to deal with this case using curved space methods in Sect. \ref{cs}, but further
work along these lines would be of interest.

\acknowledgements
We are grateful to Nick Proukakis and Stuart Coburn for discussions on the Boltzmann equation.

\bibliography{paper.bib,damping.bib}

\end{document}